\documentclass{article}[10pt]
\usepackage{spconf,amssymb,amsmath,mathrsfs}
\usepackage[dvips]{graphicx} 
\DeclareGraphicsExtensions{.bmp,.png,.pdf,.jpg}
\usepackage{amsmath}
\usepackage{times}
\usepackage{amsbsy,color}
\usepackage{latexsym}
\usepackage{amssymb}
\usepackage{graphicx}
\usepackage{subfigure}
\usepackage{amsfonts}
\usepackage{epsfig,subfigure,tikz}
\begin{document}
\title{Study of Sparsity-Aware Distributed Conjugate Gradient Algorithms for Sensor Networks}

\name{Rodrigo C. de Lamare*$^\#$ \vspace{-0.6em} }

\address{ *Department of Electronics, The University of York, England, YO10 5BB\\
{ $^\#$CETUC, Pontifical Catholic University of Rio de Janeiro, Brazil}\\
 Emails: hr648@york.ac.uk,  delamare@cetuc.puc-rio.br \vspace{-0.85em}
 \sthanks{This work was supported in part by The
University of York}}

\maketitle

\markboth{Asilomar Conference on Signals, Systems and Computers - Asilomar 2015, November $8-11^{th}$, 2015, Pacific Grove, CA} {Asilomar Conference on Signals, Systems and Computers - Asilomar 2015, November $8-11^{th}$, 2015, Pacific Grove, CA}

\begin{abstract}
This paper proposes distributed adaptive algorithms based on the conjugate
gradient (CG) method and the diffusion strategy for parameter estimation over
sensor networks. We present sparsity-aware conventional and modified
distributed CG algorithms using $l_{1}$ and log-sum penalty functions. The
proposed sparsity-aware diffusion distributed CG algorithms have an improved
performance in terms of mean square deviation (MSD) and convergence as compared
with the consensus least-mean square (Diffusion-LMS) algorithm, the diffusion
CG algorithms and a close performance to the diffusion distributed recursive
least squares (Consensus-RLS) algorithm. Numerical results show that the
proposed algorithms are reliable and can be applied in several scenarios.
\end{abstract}


\section{Introduction}

Distributed processing has become a very common and useful approach to extract
information in a network by performing estimation of the desired parameters.
The efficiency of the network depends on the communication protocol used to
exchange information between the nodes, as well as the algorithm to obtain the
parameters. Another important aspect is to prevent a failure in any agent that
may affect the operation and the performance of the network. Distributed
schemes can offer better estimation performance of the parameters as compared
with the centralized approach, based on the principle that each node
communicates with several other nodes and exploits the spatial diversity in the
networks \cite{ref1},
\cite{ref15},\cite{drr_conf},\cite{dta_conf1},\cite{dta_conf2},\cite{dta_ls}.

The main strategies for communication in distributed processing are
incremental, consensus and diffusion. In the incremental protocol, the
communication flows cyclically and the information is exchanged from one node
to the adjacent nodes. In this strategy the flow of information must be preset
at the initialization \cite{ref2}. The consensus strategy is an elegant
procedure to enforce agreement among cooperating nodes \cite{ref3}. In the
diffusion mechanism, each node communicates with the rest of the nodes
\cite{ref4} without any enforcement constraint.

In many scenarios, the parameters of unknown systems can be assumed sparse,
containing only a few large coefficients interspersed among many negligible
ones \cite{ref5}. Many studies have shown that exploiting the sparsity of a
system is beneficial to enhancing the estimation performance \cite{ref6}. Most
of the studies developed for distributed processing exploiting sparsity
\cite{delamaresp}-\cite{ref10} focus on the least-mean square (LMS) and
recursive least-squares (RLS) algorithms using different penalty functions
\cite{ref7}-\cite{ref10}. These penalty functions perform a regularization that
attracts to zero the coefficients of the parameter vector that are not
associated with the weights of interest. The most well-known and exploited
penalty functions are the $l_{0}$-norm, the
$l_{1}$-norm and the log-sum \cite{ref10}. 


The conjugate gradient (CG) algorithm has been studied and developed for
distributed processing \cite{ref12}-\cite{ref22}. The faster learning of CG
algorithms over the LMS algorithm and its lower computational complexity
combined with better numerical stability than the RLS algorithm makes it
suitable for this task. However, prior work on distributed CG techniques is
rather limited and techniques that exploit possible sparsity of the signals
have not been developed so far.

In this paper we propose distributed CG algorithms based on two variants of the
diffusion strategy for parameter estimation over sensor networks. Specifically,
we develop standard and sparsity-aware distributed CG algorithms using the
diffusion protocol and the $l_{1}$ and log-sum penalty functions. The proposed
algorithms are compared with recently reported algorithms in the literature.
The application scenario in this work is parameter estimation over sensor
networks, which can be found in many scenarios of practical interest.

This paper is organized as follows. Section II describes the system model and
the problem statement. Section III presents the proposed distributed CG
algorithm conventional and modified versions. Section IV details the proposed
sparsity-aware distributed diffusion CG algorithm. Section V presents and
discusses the simulation results. Finally, Section VI gives the conclusions.


\section{System Model and Problem Statement}


\subsection{System Model}

The network consists of N nodes that exchange information between them, where
each node represents an adaptive parameter vector with neighborhood described
by the set $N_{k}$. The main task of parameter estimation is to adjust the
unknown \textit{M}$\times 1$ weight vector $\boldsymbol\omega_{k}$ of each
node, where \textit{M} is the length of the filter \cite{ref3}. The desired
signal $d_{k,i}$ at each time \textit{i} is drawn from a random process and
given by
\begin{equation}
\ d_{k,i}=\boldsymbol\omega_{0}^{H}\boldsymbol x_{k,i}+n_{k,i}
\end{equation}
where $\boldsymbol\omega_{0}$ is the \textit{M}$\times 1$ system weight vector, $\boldsymbol x_{k,i}$ is the \textit{M}$\times 1$ input signal vector and $n_{k,i}$ is the measurement noise. The output estimate is given by
\begin{equation}
\ y_{k,i}=\boldsymbol\omega_{k,i}^{H}\boldsymbol x_{k,i}
\end{equation}
The main goal of the network is to minimize the following cost function:
\begin{equation}
\ C(\boldsymbol\omega_{k,i})=\sum_{k=1}^NE[|d_{k,i}-\boldsymbol\omega_{k,i}^{H}\boldsymbol x_{k,i}|^{2}]
\end{equation}
By solving this minimization problem it is possible to obtain the optimum solution of the weight vector at each node. The optimum solution for the cost function is given by
\begin{equation}\label{Eqn4:minimization}
\ \boldsymbol\omega_{k,i}=\boldsymbol R_{k,i}^{-1}\boldsymbol b_{k,i}
\end{equation}
where $\boldsymbol R_{k,i}=E[\boldsymbol x_{k,i}\boldsymbol x_{k,i}^{H}]$ is
the \textit{M}$\times \textit{M}$ correlation matrix of the input data vector
$\boldsymbol x_{k,i}$, and $\boldsymbol b_{k,i}=E[d_{k,i}^{*}\boldsymbol
x_{k,i} ]$ is the \textit{M}$\times 1$ cross-correlation vector between the
input data and $d_{k,i}$.

\subsection{Problem Statement}

We consider a diffusion algorithm for a network where each agent k has access
at each time instant to the realization $\{d_{k,i} , \boldsymbol x_{k,i}\}$ of
zero-mean spatial data $\{d_{k,i} , \boldsymbol x_{k,i}\}$
\cite{ref12}-\cite{dta_ls}. For a network with possibly sparse parameter
vectors, the cost function also involves a penalty function which exploits
sparsity. In this case the network needs to solve the following optimization
problem:
\begin{equation}\label{Eqn5:cost_function}
\mbox{min}\ C(\boldsymbol \omega_{k,i})=\boldsymbol \sum_{k=1}^NE[|d_{k,i}-\boldsymbol\omega_{k,i}^{H}\boldsymbol x_{k,i}|^{2}]+f(\boldsymbol \omega_{k,i})
\end{equation}
where $f(\boldsymbol\omega_{k,i})$ is a penalty function that exploits the
sparsity in the parameter vector $\boldsymbol\omega_{k,i}$. In the following
sections we focus on distributed diffusion CG algorithms to solve
(\ref{Eqn5:cost_function})

\section{Proposed Distributed Diffusion CG Algorithm}

In this section, we present the proposed distributed CG algorithm using the
diffusion strategy with a penalty function that is equal to zero. This
corresponds to the diffusion strategy without the exploitation of sparsity. We
first derive the CG algorithm and then consider the diffusion protocol.

\subsection{Derivation of the CG algorithm}

\begin{figure}[hbt]
\centering
\includegraphics[scale=0.75]{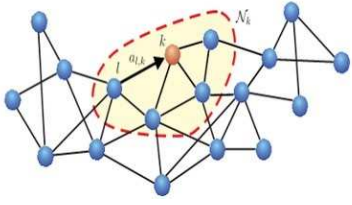}
\caption{\label{figura}Distributed consensus-based network processing.}
\label{1}
\end{figure}

The CG method can be applied to adaptive filtering problems
\cite{ref11}\cite{ref12}\cite{ref16}. The main objective in this task is to
solve (\ref{Eqn4:minimization}). The cost function for one agent is given by
\begin{equation}\label{Eqn6:Cost CG}
\ C_{CG}(\boldsymbol\omega)=\frac{1}{2}\boldsymbol\omega^{H}\boldsymbol R\omega-\boldsymbol b^{H}\boldsymbol\omega
\end{equation}
For distributed processing over sensor networks, we present the following
derivation. The CG algorithm does not need to compute the matrix inversion of
$\boldsymbol R$, which is an advantage as compared with RLS algorithms. It
computes the weights $\boldsymbol\omega_{k,i}$ for each iteration j until
convergence, i.e., $\boldsymbol\omega_{k,i}(j)$. The gradient of the method in
the negative direction is obtained as follows \cite{ref11}:
\begin{equation}\label{Eqn7:gradient method}
\ \boldsymbol g_{k,i}(j)=\boldsymbol g_{k,i}(j)-\boldsymbol R_{k,i}(j)\boldsymbol \omega_{k,i}(j)
\end{equation}
Calculating the Krylov subspace \cite{ref13} through different operations, the
recursion is given by
\begin{equation}\label{Eqn8:recursion}
\ \boldsymbol \omega_{k,i}(j)=\boldsymbol\omega_{k,i}(j-1)-\alpha(j)\boldsymbol p_{k,i}(j)
\end{equation}
where $\boldsymbol p$ is the conjugate direction vector of $\boldsymbol g$ and
$\alpha$ is the step size that minimizes the cost function in (\ref{Eqn6:Cost
CG}) by replacing (\ref{Eqn7:gradient method}) in (\ref{Eqn4:minimization}).
Both parameters are calculated as follows:
\begin{equation}\label{Eqn9:alpha}
\ \alpha(j)=\frac{\boldsymbol g_{k,i}^H(j-1)\boldsymbol g_{k,i}(j-1)}{\boldsymbol p_{k,i}^H(j)\boldsymbol R_{k,i}(j)\boldsymbol p_{k,i}(j)}
\end{equation}
\begin{equation}\label{Eqn10:direction vector}
\ \boldsymbol p_{k,i}(j)=\boldsymbol g_{k,i}(j)+\beta(j)\boldsymbol p_{k,i}(j)
\end{equation}
The parameter $\beta$ is calculated using the Gram-Schmidt orthogonalization
procedure \cite{ref14} as given by
\begin{equation}\label{Eqn11:beta}
\ \beta(j)=\frac{\boldsymbol g_{k,i}^H(j)\boldsymbol g_{k,i}(j)}{\boldsymbol g_{k,i}^H(j-1)\boldsymbol g_{k,i}(j-1)}
\end{equation}
Applying the CG method to a distributed network the cost function is expressed based on the information exchanged between all nodes $k=1,2...,N$. Each of the equations presented so far takes place at each agent during the iterations of the CG algorithm. Therefore, we have the cost function:
\begin{equation}\label{Eqn12:cost sum}
\ \ \ \ \ \ \ \ \ C_{CG}(\boldsymbol\omega_{k,i})=\frac{1}{2}\sum_{k=1}^{N}\boldsymbol\omega_{k,i}^{H}\boldsymbol R_{k,i}\boldsymbol\omega_{k,i}-\boldsymbol b_{k,i}^{H}\boldsymbol\omega_{k,i}
\end{equation}
Using the data window with an exponential decay, the resulting autocorrelation
matrix and cross-correlation vector are defined using the forgetting factor
$\lambda$ as given by
\begin{equation}\label{Eqn13:autocorrelation}
\ \boldsymbol R_{k,i}=\lambda \boldsymbol R_{k,i-1}+\boldsymbol x_{k,i}\boldsymbol x_{k,i}^{H}
\end{equation}
\begin{equation}\label{Eqn14:crosscorrelation}
\ \boldsymbol b_{k,i}=\lambda \boldsymbol b_{k,i-1}+d_{k,i}^{*}\boldsymbol x_{k,i}
\end{equation}
In the diffusion strategy, all nodes interact with their neighbors sharing and
updating the system parameter vector. Each node \textit{k} is able to run its
update simultaneously with the other agents \cite{ref1} \cite{ref4}. Figure
\ref{1} illustrates the diffusion strategy.

In diffusion protocols there are two well-known variants that switch the order
of the combination and adaptation steps, namely, Combine-then-Adapt
$\text{(CTA)}$ and Adapt-then-Combine $\text{(ATC)}$, each one based on the
connectivity among nodes. These mechanisms perform adaptation and learning at
the same time \cite{ref3}\cite{ref4}.

\subsection{CTA Diffusion Distributed CG algorithm}

In the CTA diffusion strategy, the convex combination term is first evaluated
into an intermediate state variable and subsequently used to perform the weight
update \cite{ref4}. The local estimation is given by
\begin{equation}\label{Eqn15:local estiamtion}
\ \boldsymbol\varphi_{k,i}=\sum_{l\epsilon N_{k}} a_{lk}\boldsymbol\omega_{l,i-1}
\end{equation}
where $a_{lk}$ represents the combining coefficients of the data fusion which should comply with
\begin{equation}\label{Eqn16:combinig coefficients}
\ \sum_{l\epsilon N_{k}} a_{lk}=1, l\epsilon N_{k,i}, \forall k.
\end{equation}
In this work the strategy adopted for the $a_{lk}$ combiner is the Metropolis
rule \cite{ref1} given by
\begin{equation}\label{Eqn17:Metropolis rule}
c_{kl}=\left\{\begin{array}{ll}
\frac{1}
{max\{|\mathcal{N}_k|,|\mathcal{N}_l|\}},\ \ $if\  $k\neq l$\ are linked$\\
1 - \sum\limits_{l\in \mathcal{N}_k / k} c_{kl}, \ \ $for\  $k$\ =\ $l$$,
\end{array}
\right.
\end{equation}
The distributed CTA CG algorithm based on the derivation steps obtains the
updated weight substituting (\ref{Eqn15:local estiamtion}) in
(\ref{Eqn8:recursion}), giving as result:
\begin{equation}\label{Eqn18:recursion CTA-CG}
\ \boldsymbol \omega_{k,i}(j)=\boldsymbol\omega_{k,i}(j-1)-\alpha(j)\boldsymbol p_{k,i}(j)
\end{equation}
where $\boldsymbol \omega_{k,i}(0)=\boldsymbol\varphi_{k,i}$ \cite{ref12}. The
rest of the derivation is based on the solution presented in the previous
section and the pseudo-code is detailed in Table \ref{Table1:CTA CG ALGORITHM}
\begin{table}
\caption{CTA CG ALGORITHM} \centering
\begin{tabular}{l}
\hline
Parameters initialization:\\
$\boldsymbol \omega_{k,0}=\boldsymbol 0$\\
For each time instant $i>0$ \\ 
\ \ \ \ \ \ \ \ For each agent  $k$=1,2, \ldots, N\\
\ \ \ \ \ \ \ \ \ \ \ \ \ \ \ \ $\boldsymbol R_{k,i}=\lambda \boldsymbol R_{k,i-1}+\boldsymbol x_{k,i}\boldsymbol x_{k,i}^{H}$\\
\ \ \ \ \ \ \ \ \ \ \ \ \ \ \ \ $\boldsymbol b_{k,i}=\lambda \boldsymbol b_{k,i-1}+d_{k,i}^{*}\boldsymbol x_{k,i}$\\
\ \ \ \ \ \ \ \ \ \ \ \ \ \ \ \ $\boldsymbol\varphi_{k,i}=\sum_{l\epsilon N_{k}} a_{lk}\omega_{l,i-1}$\\
\ \ \ \ \ \ \ \ \ \ \ \ \ \ \ \ $\boldsymbol \omega_{k,i}(j)=\boldsymbol\varphi_{k,i}$ \\ 
\ \ \ \ \ \ \ \ \ \ \ \ \ \ \ \ $\boldsymbol g{k,i}(0)=\boldsymbol b_{k,i}(0)-\boldsymbol R_{k,i}(0)\boldsymbol\omega_{k,i-1}$ \\
\ \ \ \ \ \ \ \ \ \ \ \ \ \ \ \ $\boldsymbol p{k,i}(0)=\boldsymbol g_{k,i}(0)$ \\
\ \ \ \ \ \ \ \ \ \ \ \ \ \ \ \ For each CG iteration $j=1$ until convergence\\
\ \ \ \ \ \ \ \ \ \ \ \ \ \ \ \ \ \ \ \ \ \ \ \ $\alpha(j)=\frac{\boldsymbol g_{k,i}^H(j-1)\boldsymbol g_{k,i}(j-1)}{\boldsymbol p_{k,i}^H(j)\boldsymbol R_{k,i}(j)\boldsymbol p_{k,i}(j)}$ \\
\ \ \ \ \ \ \ \ \ \ \ \ \ \ \ \ \ \ \ \ \ \ \ \ $\boldsymbol \omega_{k,i}(j)=\boldsymbol\omega_{k,i}(j-1)-\alpha(j)\boldsymbol p_{k,i}(j)$\\
\ \ \ \ \ \ \ \ \ \ \ \ \ \ \ \ \ \ \ \ \ \ \ \ $\boldsymbol g_{k,i}(j)=\boldsymbol g_{k,i}(j)-\alpha(j)\boldsymbol R_{k,i}(j)\boldsymbol p_{k,i}(j-1)$\\
\ \ \ \ \ \ \ \ \ \ \ \ \ \ \ \ \ \ \ \ \ \ \ \ $\beta(j)=\frac{\boldsymbol g_{k,i}^H(j)\boldsymbol g_{k,i}(j)}{\boldsymbol g_{k,i}^H(j-1)\boldsymbol g_{k,i}(j-1)}$\\
\ \ \ \ \ \ \ \ \ \ \ \ \ \ \ \ \ \ \ \ \ \ \ \ $\boldsymbol p_{k,i}(j)=\boldsymbol g_{k,i}(j)+\beta(j)\boldsymbol p_{k,i}(j-1)$\\
\ \ \ \ \ \ \ \ \ \ \ \ \ \ \ \ End For\\
\ \ \ \ \ \ \ \ \ \ \ \ \ \ \ \ $\boldsymbol\omega_{k,i} = \boldsymbol\omega_{k,i}(j_{\textit{last}})$\\
\ \ \ \ \ \ \ \ End for\\
End for\\
\hline
\end{tabular}
\label{Table1:CTA CG ALGORITHM}
\end{table}


The modified CG (MCG) algorithm comes from the conventional CG algorithm
previously presented and only requires one iteration per coefficient update.
Specifically, the residual is calculated using (\ref{Eqn7:gradient method})
(\ref{Eqn8:recursion}) and (\ref{Eqn13:autocorrelation}) \cite{ref11}:
\begin{equation}\label{Eqn19:residual CTA-MCG}
\begin{array}{ll} 
\ \boldsymbol g_{k,i}=\boldsymbol b_{k,i}-\boldsymbol R_{k,i}\boldsymbol\varphi_{k,i}\\ [2mm]
\ \ \ \ \ \ \  =\lambda\boldsymbol g_{k,i-1}-\alpha_{k,i}\boldsymbol R_{k,i}\boldsymbol p_{k,i-1}\\ [2mm]
\ \ \ \ \  +\boldsymbol x_{k,i}[d_{k,i}-\boldsymbol\omega_{k,i-1}^{H}\boldsymbol x_{k,i}]
\end{array}
\end{equation}
The previous equation (\ref{Eqn19:residual CTA-MCG}) is multiplied for the
search direction vector:
\begin{equation}\label{Eqn20:dir vector CTA-MCG}
\begin{array}{ll}
\ \boldsymbol p_{k,i}^{H} \boldsymbol g_{k,i} =\lambda\boldsymbol p_{k,i}^{H} \boldsymbol g_{k,i-1}-\alpha_{k,i}\boldsymbol R_{k,i}\boldsymbol p_{k,i-1}\\ [2mm]
\ \ \ \ \ \ \ \ \ \ \ \ \ \ \ \ \ \ \ \ \ \  +\boldsymbol p_{k,i}^{H} \boldsymbol x_{k,i}[d_{k,i}-\boldsymbol\omega_{k,i-1}^{H}\boldsymbol x_{k,i}]
\end {array}
\end{equation}
Applying the expected value, considering $\boldsymbol p_{k,i-1}$ uncorrelated
with $\boldsymbol x_{k,i}$, $d_{k,i}$ and $\boldsymbol \varphi_{k,i}$, and that
the algorithm converges the last term of (\ref{Eqn20:dir vector CTA-MCG}) can
be neglected. The line search to compute $\alpha$ has to satisfy the
convergence bound [11] given by
\begin{equation}\label{Eqn21:line search}
\begin{split}
({\lambda_f-0.5})\frac{E[\boldsymbol p_{k,i-1}^{H}\boldsymbol
g_{k,i}]}{E[\boldsymbol p_{k,i-1}^{H}\boldsymbol R_{k,i}\boldsymbol p_{k,i-1}]}
\leq E[\alpha_{k,i}] \leq \frac{E[\boldsymbol p_{k,i-1}^{H}\boldsymbol
g_{k,i}]} {E[\boldsymbol p_{k,i-1}^{H}\boldsymbol R_{k,i}\boldsymbol
p_{k,i-1}]},
\end{split}
\end{equation}\label{Eqn22:alpha modified}
\begin{equation}
\ \alpha_{k,i} = {\eta} \frac{\boldsymbol p_{k,i}^{H}\boldsymbol
g_{k,i}}{\boldsymbol p_{k,i}^{H}\boldsymbol R_{k,i} \boldsymbol p_{k,i}},
\end{equation}
where $(\lambda -0.5)\leq\eta\leq\lambda $. The Polak-Ribiere method
\cite{ref11},\cite{mcg},\cite{dcg} for the computation of $\beta$ is given by
\begin{equation}\label{Eqn23:beta modified}
\beta_{k,i} = \frac{(\boldsymbol g_{k,i}-\boldsymbol g_{k,i-1})^{H}\boldsymbol g_{k,i}}{\boldsymbol g_{k,i}^{H}\boldsymbol g_{k,i}}
\end{equation}

\subsection{ATC Distributed CG algorithm}

Similarly to CTA, the ATC protocol switches the order of the operations. The
difference lies in the variable chosen to update the weight
$\boldsymbol\omega_{k,i}$. In this case, the update estimate is the convex
combination of the adaptation step. Table \ref{Table2:ATC CG ALGORITHM} shows
the pseudo code of the ATC strategy. The MCG version for the ATC strategy is
very similar to the CTA version presented. Table \ref{Table3:ATC MCG ALGORITHM}
shows the details of the ATC MCG  algorithm taking into account the
considerations previously discussed.

\begin{table}
\caption{ATC CG ALGORITHM} \centering
\begin{tabular}{l}
\hline
Parameters initialization:\\
$\boldsymbol \omega_{k,0}=\boldsymbol 0$\\
For each time instant $i>0$ \\ 
\ \ \ \ \ \ \ \ For each agent  $k$=1,2, \ldots, N\\
\ \ \ \ \ \ \ \ \ \ \ \ \ \ \ \ $\boldsymbol R_{k,i}=\lambda \boldsymbol R_{k,i-1}+\boldsymbol x_{k,i}\boldsymbol x_{k,i}^{H}$\\
\ \ \ \ \ \ \ \ \ \ \ \ \ \ \ \ $\boldsymbol b_{k,i}=\lambda \boldsymbol b_{k,i-1}+d_{k,i}^{*}\boldsymbol x_{k,i}$\\
\ \ \ \ \ \ \ \ \ \ \ \ \ \ \ \ $\boldsymbol g{k,i}(0)=\boldsymbol b_{k,i}(0)-\boldsymbol R_{k,i}(0)\boldsymbol\omega_{k,i-1}(0)$ \\
\ \ \ \ \ \ \ \ \ \ \ \ \ \ \ \ $\boldsymbol p{k,i}(0)=\boldsymbol g_{k,i}(0)$ \\
\ \ \ \ \ \ \ \ \ \ \ \ \ \ \ \ $\boldsymbol \omega_{k,i}(0)=\boldsymbol\omega_{k,i-1}$ \\
\ \ \ \ \ \ \ \ \ \ \ \ \ \ \ \ For each CG iteration $j=1$ until convergence\\
\ \ \ \ \ \ \ \ \ \ \ \ \ \ \ \ \ \ \ \ \ \ \ \ $\alpha(j)=\frac{\boldsymbol g_{k,i}^H(j-1)\boldsymbol g_{k,i}(j-1)}{\boldsymbol p_{k,i}^H(j)\boldsymbol R_{k,i}(j)\boldsymbol p_{k,i}(j)}$ \\
\ \ \ \ \ \ \ \ \ \ \ \ \ \ \ \ \ \ \ \ \ \ \ \ $\boldsymbol \omega_{k,i}(j)=\boldsymbol\omega_{k,i}(j-1)-\alpha(j)\boldsymbol p_{k,i}(j)$\\
\ \ \ \ \ \ \ \ \ \ \ \ \ \ \ \ \ \ \ \ \ \ \ \ $\boldsymbol g_{k,i}(j)=\boldsymbol g_{k,i}(j)-\alpha(j)\boldsymbol R_{k,i}(j)\boldsymbol p_{k,i}(j-1)$\\
\ \ \ \ \ \ \ \ \ \ \ \ \ \ \ \ \ \ \ \ \ \ \ \ $\beta(j)=\frac{\boldsymbol g_{k,i}^H(j)\boldsymbol g_{k,i}(j)}{\boldsymbol g_{k,i}^H(j-1)\boldsymbol g_{k,i}(j-1)}$\\
\ \ \ \ \ \ \ \ \ \ \ \ \ \ \ \ \ \ \ \ \ \ \ \ $\boldsymbol p_{k,i}(j+1)=\boldsymbol g_{k,i}(j)+\beta(j)\boldsymbol p_{k,i}(j)$\\
\ \ \ \ \ \ \ \ \ \ \ \ \ \ \ \ End For\\
\ \ \ \ \ \ \ \ \ \ \ \ \ \ \ \ $\boldsymbol\omega_{k,i} = \sum_{l\epsilon N_{k}} a_{lk}\omega_{k,i}(j_{\textit{last}})$\\ 
\ \ \ \ \ \ \ \ End for\\
End for\\
\hline
\end{tabular}
\label{Table2:ATC CG ALGORITHM}
\end{table}

\begin{table}
\caption{ATC MCG ALGORITHM} \centering
\begin{tabular}{l}
\hline
Parameters initialization:\\
$\boldsymbol \omega_{k,0}=\boldsymbol 0$\\
For each time instant $i>0$ \\ 
\ \ \ \ \ \ \ \ For each agent  $k$=1,2, \ldots, N\\
\ \ \ \ \ \ \ \ \ \ \ \ \ \ \ \ $\boldsymbol R_{k,i}=\lambda \boldsymbol R_{k,i-1}+\boldsymbol x_{k,i}\boldsymbol x_{k,i}^{H}$\\
\ \ \ \ \ \ \ \ \ \ \ \ \ \ \ \ $\boldsymbol b_{k,i}=\lambda \boldsymbol b_{k,i-1}+d_{k,i}^{*}\boldsymbol x_{k,i}$\\
\ \ \ \ \ \ \ \ \ \ \ \ \ \ \ \ $\boldsymbol \omega_{k,i}(j)=\boldsymbol\varphi_{k,i}$ \\ 
\ \ \ \ \ \ \ \ \ \ \ \ \ \ \ \ $\boldsymbol g_{k,1}=\boldsymbol b_{k,0}$\\ 
\ \ \ \ \ \ \ \ \ \ \ \ \ \ \ \ $\boldsymbol p_{k,1}=\boldsymbol g_{k,1}$ \\
\ \ \ \ \ \ \ \ \ \ \ \ \ \ \ \ $\alpha_{k,i} = {\eta} \frac{\boldsymbol p_{k,i}^{H}\boldsymbol g_{k,i}}{\boldsymbol p_{k,i}^{H}\boldsymbol R_{k,i} \boldsymbol p_{k,i}}, (\lambda -0.5)\leq\eta\leq\lambda$ \\
\ \ \ \ \ \ \ \ \ \ \ \ \ \ \ \ $\boldsymbol\varphi_{k,i}=\boldsymbol \omega_{k,i-1}-\alpha_{k,i}\boldsymbol p_{k,i}$\\
\ \ \ \ \ \ \ \ \ \ \ \ \ \ \ \ $\boldsymbol g_{k,i}=\lambda\boldsymbol g_{k,i}-\alpha_{k,i}\boldsymbol R_{k,i}\boldsymbol p_{k,i-1}$\\                              \ \ \ \ \ \ \ \ \ \ \ \ \ \ \ \ \ \ \ \ \ \ \ \ \ \ \ \ \ $ +\boldsymbol x_{k,i}[d_{k,i}-\boldsymbol\omega_{k,i-1}^{H}\boldsymbol x_{k,i}]$\\
\ \ \ \ \ \ \ \ \ \ \ \ \ \ \ \ $\beta_{k,i} = \frac{(\boldsymbol g_{k,i}-\boldsymbol g_{k,i-1})^{H}\boldsymbol g_{k,i}}{\boldsymbol g_{k,i-1}^{H}\boldsymbol g_{k,i-1}}$\\
\ \ \ \ \ \ \ \ \ \ \ \ \ \ \ \ $\boldsymbol p_{k,i}=\boldsymbol g_{k,i}+\beta_{k,i}\boldsymbol p_{k,i-1}$\\
\ \ \ \ \ \ \ \ End for\\
\ \ \ \ \ \ \ \ $\boldsymbol \omega(i)=\sum_{l\epsilon N_{k}} a_{lk}\boldsymbol\varphi_{l,i}$\\
End for\\
\hline
\end{tabular}
\label{Table3:ATC MCG ALGORITHM}
\end{table}

\section{Proposed Sparsity-Aware Diffusion CG}

Based on the previous development of distributed CG algorithms, this section
presents the proposed distributed sparsity-aware diffusion CG algorithms using
$l_{1}$ (ZA) and \textit{log-sum} (RZA) norm penalty functions.

\subsection{ZA and RZA CG algorithms}

The cost function in this case is given by
\begin{equation}\label{Eqn24:sparse cost function}
\ \ \ \ \ \ \
C_{CG}(\boldsymbol\omega_{k,i})=\frac{1}{2}\sum_{k=1}^{N}\boldsymbol\omega_{k,i}^{H}\boldsymbol
R\omega_{k,i}-\boldsymbol b_{k,i}^{H}\boldsymbol\omega_{k,i} + f_{1},
\end{equation}
where $f_{1}$ denotes the $l_{1}$ penalty function and is defined by
\begin{equation}\label{Eqn25:function one}
\ f_{1}=\rho\|\boldsymbol\omega_{k,i}(j)\|_{1}.
\end{equation}
Applying the partial derivative of the penalty function gives
\begin{equation}\label{Eqn26:function one derivation}
\ \ \ \ \ \ \ \frac{\partial(f_{1})}{\partial(\boldsymbol\omega_{k,i}^{*})}=sgn(\boldsymbol\omega_{k,i})=\left\{\begin{array}{ll}
\frac{\boldsymbol\omega_{k,i}}
{|\boldsymbol\omega_{k,i}|},\ \ $if\  $\boldsymbol\omega_{k,i}\neq 0$$\\
0,\ \ $if\  $\boldsymbol\omega_{k,i}= 0$$,
\end{array}
\right.
\end{equation}
When the logarithmic penalty function $f_{2}$ instead $f_{1}$ is used in the
cost function, we have
\begin{equation}\label{Eqn27:function two}
\ f_{2}=\rho\sum_{i=1}^{M}\log(1+\frac{|\omega_{k,i}|}{\varepsilon}).
\end{equation}
The partial derivative of the penalty function applied with respect to
$\boldsymbol\omega_{k,i}^{*}$ is described by
\begin{equation}\label{Eqn28:function two derivation}
\ \ \ \ \ \ \ \frac{\partial(f_{1})}{\partial(\boldsymbol\omega_{k,i}^{*})}=\frac{sgn(\boldsymbol\omega_{k,i})}{1+\varepsilon\|\boldsymbol\omega_{k,i}\|_{1}}
\end{equation}
In both cases these sparsity-aware algorithms attract to zero the values of the
parameter vector which are very small or are not useful. This results in an
algorithm with a faster convergence and lower MSD values as can be seen in the
following sections. Using the penalty functions (\ref{Eqn26:function one
derivation}) and (\ref{Eqn28:function two derivation}), we obtain the
sparsity-aware algorithms with the CTA and ATC strategies. Table
\ref{Table4:SPARSE-AWARE ATC CG ALGORITHM} shows the sparsity-aware method for
the ACT protocol. In case of CTA, the same steps applied with the ZA or RZA
penalty functions to the steps previously presented in Section III are carried
out.
\begin{table}
\caption{Sparsity-aware ATC CG Algorithm} \centering
\begin{tabular}{l}
\hline
Parameters initialization:\\
$\boldsymbol \omega_{k,0}=\boldsymbol 0$\\
For each time instant $i>0$ \\ 
\ \ \ \ \ \ \ \ For each agent  $k$=1,2, \ldots, N\\
\ \ \ \ \ \ \ \ \ \ \ \ \ \ \ \ $\boldsymbol R_{k,i}=\lambda \boldsymbol R_{k,i-1}+\boldsymbol x_{k,i}\boldsymbol x_{k,i}^{H}$\\
\ \ \ \ \ \ \ \ \ \ \ \ \ \ \ \ $\boldsymbol b_{k,i}=\lambda \boldsymbol b_{k,i-1}+d_{k,i}^{*}\boldsymbol x_{k,i}$\\
\ \ \ \ \ \ \ \ \ \ \ \ \ \ \ \ $\boldsymbol g{k,i}(0)=\boldsymbol b_{k,i}(0)-\boldsymbol R_{k,i}(0)\boldsymbol\omega_{k,i-1}(0)$ \\
\ \ \ \ \ \ \ \ \ \ \ \ \ \ \ \ $\boldsymbol p{k,i}(0)=\boldsymbol g_{k,i}(0)$ \\
\ \ \ \ \ \ \ \ \ \ \ \ \ \ \ \ $\boldsymbol \omega_{k,i}(0)=\boldsymbol\omega_{k,i-1}$ \\
\ \ \ \ \ \ \ \ \ \ \ \ \ \ \ \ For each CG iteration $j=1$ until convergence\\
\ \ \ \ \ \ \ \ \ \ \ \ \ \ \ \ \ \ \ \ \ \ \ \ $\alpha(j)=\frac{\boldsymbol g_{k,i}^H(j-1)\boldsymbol g_{k,i}(j-1)}{\boldsymbol p_{k,i}^H(j)\boldsymbol R_{k,i}(j)\boldsymbol p_{k,i}(j)}$ \\
\ \ \ \ \ \ \ \ \ \ \ \ \ \ \ \ \ \ \ \ \ \ \ \ $\boldsymbol \omega_{k,i}(j)=\boldsymbol\omega_{k,i}(j-1)-\alpha(j)\boldsymbol p_{k,i}(j)$\\
\ \ \ \ \ \ \ \ \ \ \ \ \ \ \ \ \ \ \ \ \ \ \ \ $\boldsymbol g_{k,i}(j)=\boldsymbol g_{k,i}(j)-\alpha(j)\boldsymbol R_{k,i}(j)\boldsymbol p_{k,i}(j-1)$\\
\ \ \ \ \ \ \ \ \ \ \ \ \ \ \ \ \ \ \ \ \ \ \ \ $\beta(j)=\frac{\boldsymbol g_{k,i}^H(j)\boldsymbol g_{k,i}(j)}{\boldsymbol g_{k,i}^H(j-1)\boldsymbol g_{k,i}(j-1)}$\\
\ \ \ \ \ \ \ \ \ \ \ \ \ \ \ \ \ \ \ \ \ \ \ \ $\boldsymbol p_{k,i}(j+1)=\boldsymbol g_{k,i}(j)+\beta(j)\boldsymbol p_{k,i}(j)$\\
\ \ \ \ \ \ \ \ \ \ \ \ \ \ \ \ End For\\
\ \ \ \ \ \ \ \ \ \ \ \ \ \ \ \ $\boldsymbol\omega_{k,i} = \sum_{l\epsilon N_{k}} a_{lk}\omega_{k,i}(j_{\textit{last}})- \rho\frac{\partial(f_{1,2})}{\partial\boldsymbol \omega_{k,i}^{*}}$\\ 
\ \ \ \ \ \ \ \ End for\\
End for\\
\hline
\end{tabular}
\label{Table4:SPARSE-AWARE ATC CG ALGORITHM}
\end{table}

\subsection{ZA and RZA Modified CG algorithms}

The ATC and CTA MCG algorithms are very similar as presented in previous
section, including the penalty functions. In the ATC strategy, the generation
of the first state resulting from the adaptation step, is used in the final
update.

\subsection{Computational Complexity}

The Table \ref{Table5:COMPUTATIONAL COMPLEXITY of DIFFUSION CG METHODS} shows
the computational complexity of all diffusion distributed methods proposed in
terms of additions and multiplications.

\begin{table}[htb]
\caption{Computational Complexity of diffusion CG algorithms}
\begin{center}
\begin{tabular}{|c|c|c| }
\hline
Method&Additions&Multiplications\\
\hline
CTA-CG&$L(M^{2}+2M)$&$L(2M^{2}+4M)$\\
&$+LJ(2M^{2}+6M-3)$ &$+LJ(3M^{2}+4M-1)$\\
\hline
ATC-CG&$L(M^{2}+3M-1)$&$L(2M^{2}+3M)$\\
&$+LJ(M^{2}+6M-3)$ &$+LJ(3M^{2}+4M-1)$\\
\hline
CTA-MCG&$L(3M^{2}+9M-4)$&$L(4M^{2}+9M-1)$\\
\hline
ATC-MCG &$L(4M^{2}+9M-3)$&$L(6M^{2}+8M-1)$\\
\hline
ZA-CTA-CG &$L(M^{2}+3M)$&$L(2M^{2}+5M)$\\
&$+LJ(2M^{2}+6M-3)$ &$+LJ(3M^{2}+4M-1)$\\
\hline
ZA-ATC-CG &$L(M^{2}+3M)$&$L(2M^{2}+5M)$\\
&$+LJ(2M^{2}+6M-3)$ &$+LJ(3M^{2}+4M-1)$\\
\hline
ZA-CTA-MCG &$L(3M^{2}+10M-4)$&$(4M^{2}+10M-1)$\\
\hline
ZA-ATC-MCG &$L(4M^{2}+10M-3)$&$(6M^{2}+9M-1)$\\
\hline
RZA-CTA-CG &$L(M^{2}+2M)$&$L(2M^{2}+4M)$\\
&$+LJ(2M^{2}+8M-3)$&$+ LJ(3M^{2}+6M-1)$\\
\hline
RZA-ATC-CG &$L(M^{2}+3M-1)$&$L(2M^{2}+3M)$\\
&$+LJ(2M^{2}+8M-3)$&$+ LJ(3M^{2}+6M-1)$\\
\hline
RZA-CTA-MCG &$L(3M^{2}+11M-4)$&$L(4M^{2}+11M-1)$\\
\hline
RZA-ATC-MCG &$L(4M^{2}+11M-3)$&$L(6M^{2}+10M-1)$\\
\hline
\end{tabular}
\end{center}
\label{Table5:COMPUTATIONAL COMPLEXITY of DIFFUSION CG METHODS}
\end{table}
It can be seen that the complexity of the modified versions is lower than the conventional methods

\section{Simulations Results}

In this section, we evaluated the proposed distributed diffusion CG algorithms
and compare them with existing algorithms. The results are based on the mean
square deviation MSD of the network. We consider a network with $20$ nodes and
$1000$ iterations per run. Each iteration corresponds to a time instant. The
results are averaged over $100$ experiments. The length of the filter is $10$
and the variance of the input signal 1, which has been modeled as a complex
Gaussian noise and the SNR is $30$ dB.

\subsection{Comparison between standard and sparsity-aware CTA distributed CG algorithms}

For the standard CTA CG version, the system parameter vector was randomly set.
In the case of the sparsity-aware algorithms it was set to two values equal to
one and the remaining values were set to zero. After all the iterations, the
performance of each algorithm in terms of MSD is shown in Fig. \ref{2}. The
results show that the sparsity-aware versions outperforms the standard versions
and the best results are obtained for the RZA versions. At the same time the
MCG algorithms have a better performance than the standard ones.


\subsection{Comparison between sparsity-aware ATC distributed CG algorithms.}

The same configuration used before was set for CTA strategy. Fig. \ref{3} below
shows the performance of the results in the simulations. Fig. \ref{4} shows the
comparison between the consensus and diffusion algorithms with RZA. It can be
observed that the diffusion ATC CG algorithm has a faster convergence as
compared to the CTA and consensus strategy, as well as the MSD value at steady
state.

%
%
%

\section{Conclusions}

In this work we proposed distributed CG algorithms for parameter estimation
over sensor networks as well as the modified versions of them. The proposed ATC
diffusion CG algorithms have a faster convergence than the CTA. The ATC
strategy outperforms both consensus[ref] and CTA protocols. In all cases, the
modified versions obtained the low MSD values and faster convergence rate.
Simulations have shown that the proposed distributed CG algorithms are suitable
techniques for adaptive parameter estimation problems.







\end{document}